\documentstyle [12pt,aaspp4]{article}
\def\beq{\begin{equation}}
\def\eeq{\end{equation}}
\def\bey{\begin{eqnarray}}
\def\eey{\end{eqnarray}}
\def\pppm{\rm P^3M}
\def\mpc{\,h^{-1}{\rm {Mpc}}}
\def\mpci{\,h {\rm {Mpc}}^{-1}}
\def\kms{\,{\rm {km\, s^{-1}}}}

\def\gs{\mathrel{\raise1.16pt\hbox{$>$}\kern-7.0pt
\lower3.06pt\hbox{{$\scriptstyle \sim$}}}}
\def\ls{\mathrel{\raise1.16pt\hbox{$<$}\kern-7.0pt
\lower3.06pt\hbox{{$\scriptstyle \sim$}}}}
\def\gtsima{$\; \buildrel > \over \sim \;$}
\def\ltsima{$\; \buildrel < \over \sim \;$}
\def\prosima{$\; \buildrel \propto \over \sim \;$}
\def\gsim{\lower.5ex\hbox{\gtsima}}
\def\lsim{\lower.5ex\hbox{\ltsima}}
\def\simgt{\lower.5ex\hbox{\gtsima}}
\def\simlt{\lower.5ex\hbox{\ltsima}}
\def\simpr{\lower.5ex\hbox{\prosima}}

\begin{document}
\title {
Scaling properties of the redshift power spectrum: theoretical models}
\author {Y.P. Jing$^{1,2,3,4}$, G. B\"orner$^{1,4,5}$} 
\affil{$ ^1$
Shanghai Astronomical Observatory, the Partner Group of MPI f\"ur
Astrophysik, Nandan Road 80, Shanghai 200030, China}
\affil {$^2$ National Astronomical Observatories, Chinese Academy of
Sciences, Beijing 100012, China}
\affil{$ ^3$ National Astronomical Observatory, Mitaka,
 Tokyo 181-8588, Japan}
\affil{$ ^4$
Research Center for the Early Universe, School of Science, University
of Tokyo, Bunkyo-ku, Tokyo 113, Japan}
\affil {$ ^5$
Max-Planck-Institut f\"ur Astrophysik, Karl-Schwarzschild-Strasse 1,
85748 Garching, Germany}
\affil {e-mail: ypjing@center.shao.ac.cn, ~grb@mpa-garching.mpg.de}
\received{---------------}
\accepted{---------------}

\begin{abstract}

We report the results of an analysis of the redshift power spectrum
$P^S(k,\mu)$ in three typical Cold Dark Matter (CDM) cosmological
models, where $\mu$ is the cosine of the angle between the wave vector
and the line--of--sight. Two distinct biased tracers derived from the
primordial density peaks of Bardeen et al. and the
cluster--underweight model of Jing, Mo, \& B\"orner are considered in
addition to the pure dark matter models. Based on a large set of high
resolution simulations, we have measured the redshift power spectrum
for the three tracers from the linear to the nonlinear regime.  We
investigate the validity of the relation --guessed from linear
theory--in the nonlinear regime
\beq
P^S(k,\mu)=P^R(k)[1+\beta\mu^2]^2D(k,\mu,\sigma_{12}(k))\,,
\eeq
where $P^R(k)$ is the real space power spectrum, and $\beta$ equals
$\Omega_0^{0.6}/b_l$. The damping function $D$ which should generally
depend on $k$, $\mu$, and $\sigma_{12}(k)$, is found to be a function
of only one variable $k\mu\sigma_{12}(k)$.  This scaling behavior
extends into the nonlinear regime, while $D$ can be accurately
expressed as a Lorentz function-- well known from linear theory-- for
values $D > 0.1$. The difference between $\sigma_{12}(k)$ and the
pairwise velocity dispersion defined by the 3--D peculiar velocity of
the simulations (taking $r=1/k$) is about $15\%$. Therefore
$\sigma_{12}(k)$ is a good indicator of the pairwise velocity
dispersion. The exact functional form of $D$ depends on the
cosmological model and on the bias scheme. We have given an accurate
fitting formula for the functional form of $D$ for the models studied.

\end{abstract}


\section {Introduction}

In general the redshift of a galaxy gives a reasonable measure of its
distance. It is, however, exact only when the galaxy follows the
linear Hubble flow, and there are two distinct deviations from that:
observations of high redshift objects require an assumption about the
cosmic geometry, and choosing, e.g., a wrong cosmological model may
create for an intrinsically isotropic spatial distribution of objects
anisotropies along the line of sight and along the line projected
perpendicularly
(\cite{1979Natur.281..358A}; \cite{1996ApJ...470L...1M}; 
\cite{1996MNRAS.282..877B}).
The peculiar motions of galaxies induced by the gravitational field of
clumpy structures also change the true distribution
(\cite{1973ApJ...184..329G}; \cite{1976ApJ...208...20T}; 
\cite{1983ApJ...267..465D}; \cite{1987MNRAS.227....1K}).
Therefore in redshift space we find a distorted picture of the spatial
distribution. This may seem unfortunate, but, on the other hand,
it allows to estimate the statistics of the clustering process.
A careful modeling of these effects can yield valuable
determinations of the cosmological model parameters, the power
spectrum, and the bias of these objects.

The statistic widely used for measuring the redshift distortion is the
redshift two-point correlation function or equivalently its Fourier
counterpart, the redshift power spectrum. In this paper we will focus
our discussion on the redshift power spectrum. The models for the
cosmological geometry effect on these statistics have been well
established (\cite{1996ApJ...470L...1M}; \cite{1996MNRAS.282..877B}),
since there exists a simple mathematical mapping of a redshift spatial
distribution from one cosmological model to another. More
uncertainties however exist in the theoretical modeling of the effects
of the peculiar motion.  These uncertainties will affect the modeling
of the redshift distortion not only at low redshift but also at high
redshift, since the peculiar motion is likely to be also important for
redshift surveys of quasars and galaxies
(e.g. \cite{adelbergeretal1998}; \cite{steideletal1997}) at high
redshift (\cite{1999PThPS.133..183S}; \cite{2000ApJ...528...30M}).
Therefore it is highly desirable to have an accurate model for the
peculiar motion effect.

In the limits of a linear density perturbation and of a linear galaxy
bias, the redshift power spectrum $P^S_{\rm l}(k,\mu)$ can be
accurately derived (\cite{1987MNRAS.227....1K}),
\beq
P^S_{\rm l}(k,\mu) = P^R_{\rm l}(k)[1+\beta \mu^2]^2\,.
\eeq
In the above equation, $\mu$ is the cosine of the angle between the
wavevector $\bf k$ and the line-of-sight; $P^R_{\rm l}$ is the linear
power spectrum in  real space; $\beta$ equals $\Omega_0^{0.6}/b_l$, where
$b_l$ is the linear bias parameter and $\Omega_0$ is the density
parameter of the universe. Throughout the paper, we will use
superscripts $S$ and $R$ to denote  quantities in  redshift space and
in real space respectively.  In  another extreme limit of collapsed
objects (the finger-of-God effect), the redshift distortion was found
{\it observationally} to be described by an exponential
distribution function (DF) for the pairwise velocity
(\cite{1976Ap&SS..45....3P}; \cite{1983ApJ...267..465D}), though significant
uncertainties must be allowed for these observations. Subsequent
theoretical studies based on  numerical simulations
(\cite{1988MNRAS.235..715E}), the \cite{1974ApJ...187..425P} theory
(\cite{1996ApJ...467...19D}; \cite{1996MNRAS.279.1310S}) and the Zeldovich
approximations (\cite{1998ApJ...492..421S}) have confirmed that the
distribution function of the pairwise velocity can be well
approximated by an exponential form for the {\it dark matter} in
currently favored CDM models and in some scale-free hierarchical
clustering models.  Based on the assumptions that a) the DF of the
pairwise velocity has an exponential form, b) the linear
(\cite{1987MNRAS.227....1K}) and the nonlinear (Finger-of-God) effects are
separable, and c) there is only weak coupling between the density and
the non-linear motion, it is not difficult to derive an {\it ansatz}
for the redshift distortion of the power spectrum on {\it all} scales
(\cite{1994MNRAS.267.1020P}; \cite{1995MNRAS.275..515C}):
\beq 
P^S(k,\mu)=P^R(k)[1+\beta
\mu^2]^2{1\over 1+{1\over 2}(k\mu\sigma_v)^2}\,.  
\label{lorentz}
\eeq 
This formula has been compared to N-body simulations
(\cite{1994MNRAS.267..785C}; \cite{1995MNRAS.275..515C};
\cite{1997ApJ...475..414B};
\cite{2000ApJ...528...30M}), and it has turned out that  
equation (\ref{lorentz}) describes the
redshift power spectrum of dark matter accurately on large scales
($k\ls 0.2 \mpci$, i.e. in the linear and quasi-linear regime) with
$\sigma_v$ equal to the pairwise velocity dispersion at large
separation (\cite{2000ApJ...528...30M}).

Equation (\ref{lorentz}) can at best be an approximation for
describing the redshift power spectrum for several reasons. For the
clustering on large scales, the coupling between the non-linear motion
and the structures is weak and the DF of the pairwise velocity is
described well by the exponential form (at least in many hierarchical
clustering models). This might be the reason why equation
(\ref{lorentz}) has been found to be in good agreement with numerical
simulations on large enough scales. However, at smaller and smaller
scales, the coupling between the non-linear motions and the structures
becomes stronger and stronger. It is also probably not a valid
procedure to extrapolate the model of linear motion
(\cite{1987MNRAS.227....1K}) to the non-linear regime to model the
infall on small scales.  Furthermore it has been found in simulations
that the DF of the pairwise velocity is significantly skewed in the
quasi-linear regime to the approaching velocity of particle pairs
(\cite{1998ApJ...504L...1J}; \cite{2000ApJ...528...30M}). For all these
reasons, we may expect that equation (\ref{lorentz}) breaks down on
non-linear scales. Some deviation of this model from simulation
data could already be seen in previous studies
(\cite{1995MNRAS.275..515C}; \cite{1997ApJ...475..414B};
\cite{2000ApJ...528...30M}) even in the quasi-linear regime ($k\gs
0.5\mpci$), though the focus of those studies is on the agreement of
the equation with the simulation data in the linear regime.

We devote this paper to studying the redshift power spectrum in the
non-linear and quasi-linear regimes. With the help of a large set of
high resolution simulations, we measure the redshift power spectrum
for dark matter from linear to strongly non-linear regimes. We find
that while equation (\ref{lorentz}) starts to break down in the
quasi-linear regime, there exists a scaling relation of the variable
$k\mu\sigma_v$ for the non-linear effects (motion and coupling between
density and velocity). The existence of this scaling relation is not
trivial, since there is a strong coupling between the velocity and the
density on non-linear scales. But such a scaling relation would be
very useful for studying and quantifying the redshift power spectrum
on small scales in observational catalogs (\S 4). Since the galaxies
in the universe do not generally trace the distribution of the
underlying matter, we will also study the redshift power spectrum for
two plausible bias models: the primordial peaks
(\cite{1986ApJ...304...15B}) and the cluster-under-weight bias
(\cite{1998ApJ...494....1J}; hereafter JMB98). We find that scaling
relations of the variable $k\mu\sigma_v$ exist in both bias
models. The scaling form as a function of the variable $k\mu\sigma_v$
depends on the bias model.  Therefore a determination of this scaling
function from observations may be used to discriminate between
different models of galaxy formation.

The paper is arranged as follows: in the next section, we will
describe the simulation samples and the bias models used.  The
techniques for measuring the redshift power spectrum are outlined, and
the results are presented in section 3. The final section, \S 4, is
devoted to discussion and conclusions.
 
\section{Simulation sample and simple bias models}

We study the redshift distortion of the power spectrum for three Cold
Dark Matter (CDM) models, i.e. the standard CDM (SCDM), a flat
low-density CDM (LCDM), and an open CDM (OCDM). The model parameters
are given in Table 1, with $\Omega_0$ for the density parameter, with
$\lambda_0$ for the cosmological constant, and with $\Gamma=\Omega_0
h$ and $\sigma_8$ respectively for the shape parameter and the
normalization of the linear power spectrum. These models constitute a
typical set of CDM models. Our N-body simulations are those of
\cite{1998ApJ...494L...5J} and \cite{jing1998}, with  box sizes of $100
\mpc$ and $300 \mpc$.  Each model has three to four statistical
realizations for one box size. All the simulations employ $256^3$
($\approx 17 $ million) particles and were generated with our $\pppm$
code on the Fujitsu vector machines at the National Astronomical
Observatory of Japan. The simulations have been used for studying a
number of cosmological problems. A summary of these applications and
the simulation details can be found in \cite{jing2000}.

While we can extract valuable information on the dark matter
distribution from N-body simulations, galaxies observed in the
Universe most likely do not exactly trace the underlying dark matter
distribution. The so-called ``bias'' of the galaxy distribution with
respect to the underlying matter distribution remains one of the
unsolved outstanding problems in cosmology. To study the effect of the
bias on the redshift distortion, we take two simple bias models: the
primordial peak model of Bardeen et al. (1986) and the
cluster-under-weighted (CLW) model of JMB98. Because of the
long-wavelength modulation of the primordial density perturbation, the
peak model ascribes more weight to the high density (cluster) regions
in the spatial distribution than the pure dark matter model, while the
CLW model, by its construction, gives less weight to the cluster
regions than the dark matter model. The redshift distortion in the
non-linear regime is known to be sensitive to the weighting of the
cluster regions. These bias models cover many facets of the bias
effect on the redshift distortion.

In our peak model, peaks are defined as fluctuations with more than 2
times the rms fluctuation of the primordial density field smoothed
with a Gaussian window $\exp(-r^2/2r_s^2)$. The window width is taken
to be $r_s=0.54\mpc$ so that the peaks are relevant for galactic-sized
objects.  We follow the prescription of
\cite{1987ApJ...313..505W} to assign an expectation number of
peaks to each simulation particle in  Lagrangian space (see
\cite{1994A&A...284..703J} for a detailed description of our
algorithm).  During the dynamical evolution, these peaks stay with
the particles to which they are assigned. Because our simulations have
a high mass resolution, the expectation number of peaks per particle
is always less than 1.

Our cluster weighting bias model is the same as that of
JMB98. Specifically, the number of galaxies per unit dark matter mass
$N/M$ is proportional to $M^{-0.08}$ within a massive halo of mass
$M$, i.e. the cluster regions are under-weighted.  We identify
clusters in the N-body simulations using the friends-of-friends method
with a linkage parameter equal to 0.2 times the mean separation of
particles.  We randomly throw away particles in clusters according to
the above bias model. Typically 10 percent of the total number of
simulation particles, mostly in cluster regions, have been left out.
This phenomenological model was proposed by JMB98 to reconcile the CDM
models with their measurement for the two-point correlation function
and the pairwise velocity dispersion in the Las Campanas Redshift
Survey. This empirical model has also received support from the
observation of the CNOC clusters (\cite{1996ApJ...462...32C}) where a
similar trend of $N/M$ has shown up, though the scatter in the
observation is still large.  There is evidence that this empirical
model is also consistent with semi-analytical models of galaxy
formation which incorporate star formation (\cite{bensonetal1999}),
since these semi-analytical models have produced predictions for the
two-point correlation function and the velocity dispersion quite
similar to the empirical model.

The bias factors $b(k)$ of these biased models, which are measured
from the square root of the ratio of the real space power spectrum of
the biased tracer to that of the underlying dark matter, are presented
in Figure~\ref{fig1}. The bias function is a constant $b_l$ on large
scales ($k<0.08 \mpci$). It rises on small scales in the peak models,
but slightly falls with $k$ in the cluster-weight models (though it
does not rise or fall monotonically at the smallest scales). The
effects of the non-linear bias functions on the linear Kaiser effect
are examined by looking at the ratio
$[1+\mu^2\Omega_0^{0.6}/b(k)]^2/[1+\mu^2\Omega_0^{0.6}/b_l]^2$. For
each $k$, the ratio reaches the maximum deviation from $1$ at
$\mu=1$. Figure~\ref{fig2} shows that replacing $b(k)$ with $b_l$ in
modeling the linear Kaiser effect could result in an error of $\ls
20\%$ in the damping factor (\S 3) in the non-linear regime. Since the
linear bias can be analytically calculated for the bias models
(e.g. \cite{matsubara_bias}) and an error of $\sim 20\%$ 
in $D$ is tolerable in this study (cf. Fig.~\ref{fig4} 
and Fig.~\ref{fig5}), we will use the
linear bias factor to model the linear Kaiser effect throughout this
paper.

\section{Redshift power spectrum}

\subsection{Measurement method}
We choose the third axis as the line-of-sight direction, and map the
coordinate positions of the simulation particles from real space to
redshift space by taking into account each particle's peculiar
velocity. Periodic boundary conditions are used to place back into the
simulation box those particles which are outside of the box in
redshift space. A grid of $640^3$ uniformly spaced points is placed
within the simulation box. The Nearest-Grid-Point (NGP) method is used
to get the density of dark matter or peaks on the grid. This grid of
density is transformed to the density distribution in Fourier space
$\delta({\bf k})$ using the Fast Fourier Transform (FFT) method. The
Nyquist wavenumber $k_N$ is about $6.7\mpci$ for a box size of
$300\mpc$ and $20 \mpci$ for a box size of $100\mpc$. We take linear
bins for $\mu$ with $\Delta \mu =0.1$, with an additional bin at
$\mu=0$. For the wavenumber $k$, equal logarithmic bins $\Delta
\lg(k)=0.1$ are taken from $0.05k_N$ to $k_N/4$. The lower limit for
$k$ is chosen such that there are sufficient modes and the
sample-to-sample fluctuation in $\delta({\bf k})$ is small. The upper
limit is taken such that the biases introduced by the FFT method are
negligible.

The assignment of mass and peaks to a grid for FFT brings about
artificial smearing as well as artificial anisotropy to the density
field $\delta({\bf k})$ near the Nyquist wavenumber. These artificial
effects on the power spectrum measurement have been discussed in
detail by \cite{jing1992}. Although these effects might be corrected
with an iterative procedure(cf Jing 2000), we adopt a simpler approach
here. We limit our discussion to small wavenumbers, where these
effects become negligible. Indeed, for the NGP assignment scheme
adopted here, these effects are seen to diminish for wavenumbers $k
\le k_N/4$, according to Jing (1992, 2000). To show this point
quantitatively, we measure the redshift power spectrum for one LCDM
simulation of box size $100\mpc$ with $320^3$ and $640^3$ grid points
respectively. The ratio of these redshift power spectra at $k=2.3
\mpci$, about one fourth of the Nyquist wavenumber for the grid points
of $320^3$, is 0.98 and nearly independent of $\mu$. Thus the redshift power
spectrum is underestimated by only 2 percent in the case of
$320^3$. This accuracy is good enough for this study. In the
following, our results will be presented for $k\le k_N/4$.

\subsection{Results}

The redshift power spectrum $P^S(k,\mu)$ which we have measured for
the dark matter is presented in the top panels of Figure 3. Different
symbols are used for $P^S(k,\mu)$ at different $k$, with $k$
increasing from the symbols at the top to those at the bottom. The
smallest and the largest values of $k$ are $0.35\mpci$ and $3.4\mpci$
respectively, and the increment of $k$ between two successive sets of
symbols is $\Delta
\log_{10}k=0.2$ approximately. Thus we obtain a sequence for
the redshift power spectrum from the quasi-linear to the highly
non-linear regimes. The non-linear peculiar motion suppresses the
clustering along the line-of-sight, with the effect manifested more
prominently at larger $k$ and larger $\mu$. The power spectrum is
suppressed by 3 magnitudes along the line-of-sight at the largest $k$ values,
indicating that we have approached the highly non-linear
regime. In all three dark matter models, we find these same
qualitative features for $P^S(k,\mu)$ .

The curves for each $k$, which always start from the data points at
$\mu=0$, are the model prediction of Equation (\ref{lorentz}) with
$\beta=\Omega_0^{0.6}$ for each simulation. For $\sigma_v(k)$, we have
used the pairwise velocity dispersion at the separation $r=1/k$
\footnote{When $\sigma_v(k)$ and the pairwise velocity dispersion 
$v_{PVD}(r)$ are compared, we use $r=1/k$ in this paper instead of the
{\it correct} relation $r=2\pi/k$. The reason is purely because
$\sigma_v(k)$ and $v_{PVD}(r)$ are better matched when $r=1/k$ is used
(see Figure~\ref{fig6}).}.  The model agrees well with the simulation
data for large scales $k\ls 0.5\mpci$. But at large $k$ where the
non-linearity is strong, the model predicts a significantly slower
decrease with $\mu$ than the simulation data. Adjusting the value for
$\sigma_v(k)$ could not produce a better match between the model and
the simulation data.

The redshift power spectra for the peaks and the cluster-weighted
particles are shown in the middle and bottom panels of Figure 3
respectively. The qualitative features of these biased models are very
similar to those of the dark matter. But with a closer look at the
figures, we can easily find that among the three models, the peak
model has the strongest dependence on $\mu$ and the cluster weighted
model has the weakest. The results are expected, since the cluster
regions are over-- weighted in the peak model and under-- weighted in
the CLW model relative to the pure dark matter model. When calculating
predictions for these biased models, we use the bias parameter $b_l$
determined on the linear scale. The simulation results of the biased
models, like those of the pure dark matter models, show a faster
decrease with $\mu$ for high $k$, indicating that equation
(\ref{lorentz}) is not adequate for describing the redshift distortion
in the highly non-linear regime even if we let $\sigma_v$ be a
function of $k$.

It is not unexpected that equation (\ref{lorentz}) breaks down in the
non-linear regime for the reasons outlined in Section 1. In order to
study the non-linear behavior of the redshift power spectrum, we
examine the relation
\beq D(k,\mu) \equiv
{P^S(k,\mu)\over P^R(k)(1+\beta \mu^2)^2}\,,
\label{dd}
\eeq 
where we take the damping function now to be an unknown quantity
$D(k,\mu)$ which should be determined from the known expressions on
the right hand side.  The factor $(1+\beta \mu^2)^2$ accounts for the
linear distortion of the power spectrum. The power spectrum $P^R(k)$
in real space is measured as in Jing (2000). This will be used for the
denominator of equation (\ref{dd}). In equation(\ref{lorentz}), the
damping function has the Lorentz form ,i.e. it is a function of
$k\mu\sigma_v$ only.  Despite the fact that the Lorentz form is
inadequate in the non-linear regime, we find that the damping function
is approximately a function of $k\mu\sigma_v$. We take $800\kms$ for
the value of $\sigma_v$ which is close to the simulation value. The
results are plotted in Figure 4 for the CDM and different bias models,
with different symbols for different wavenumbers $k$ (as in Figure 3).
The points for different values of $k$ fall on top of each other, and
this demonstrates that the damping function is indeed approximately a
function of $k\mu\sigma_v$.  Furthermore the damping function in all
the different models falls more steeply than the Lorentz form when
$D(k,\mu)< 0.1$, which is consistent with Figure 3.

Although the damping function at different $k$ and $\mu$ is {\it
approximately} a scaling function of $k\mu\sigma_v$, there exist
small but significant systematic scatters of $D(k,\mu)$ for different
$k$ along the horizontal axis $k\mu\sigma_v$. The shifts amount
typically to a few tens percent in $k\mu\sigma_v$. The reason could be
that the velocity dispersion is not a constant. In fact, it is well
known that the pairwise velocity dispersion in coordinate space is a
function of the pairwise separation $r$ and in the CDM models the
pairwise velocity dispersion peaks at $r\approx 2\mpc$
(e.g. JMB98). Therefore, we relax the assumption of $\sigma_v={\rm
constant}$ and let $\sigma_v$ vary with $k$.

Figure 5 shows the damping function in the different models as a
function of $k\mu\sigma_v$, where $\sigma_v$ is allowed to vary with
$k$. We have taken values for $\sigma_v(k)$ such that the damping
function $D(k\mu\sigma_v)$ of two neighboring $k$ bins matches
best. This determines the relative values of $\sigma_v(k)$.  In order
to fix the absolute values, we fit the damping function with the
Lorentz form for $D(k,\mu) > 0.1$. The Lorentz form for $D(k,\mu) >
0.1$ describes well the simulation data of all the models (see more
discussion below). The values of $\sigma_v (k)$ determined in this way
are plotted in Figure 6.

The damping function $D(k,\mu)$, plotted as a function of
$k\mu\sigma_v$, has a much smaller scatter with a variable $\sigma_v$
than with a constant one. Although it is not surprising that taking a
variable $\sigma_v(k)$ improves the scaling relation of $D(k,\mu)$
(since it gives more freedom than a constant $\sigma_v$), it is far
from trivial to have such a good scaling relation with a variable
$\sigma_v(k)$, since $k\mu\sigma_v$ can vary by an order of magnitude
even for a single $k$ and we do not adjust $D(k,\mu)$.

Moreover, the qualitative behavior of $\sigma_v(k)$ is very similar to
that of the pairwise velocity dispersion (PVD) $\sigma_{PVD}(r)$ which
is measured directly from the 3-dimensional peculiar velocity in the
simulation.  $\sigma_v(k)$ peaks at $k\approx 1\mpc$ and gently falls
when $k$ increases or decreases. This compares well with the pairwise
velocity dispersion $\sigma_{PVD}(r)$ (curves in Figure 6), where we
have arbitrarily assumed $k=1/r$ for the horizontal axis.  In fact, we
can use $\sigma_{PVD}$ from the simulation data for the variable
$\sigma(k)$ and obtain damping functions very similar to the graphs
shown in Figure 5.  Small differences between the use of
$\sigma_{PVD}(r)$ and of $\sigma_v(k)$ are expected, since there is no
reason that they should be exactly the same. But the fact that the
difference between $\sigma_{PVD}(r)$ and $\sigma_v(k)$ is less than
20\% (if $k\propto 1/r$ is used) is very encouraging: $\sigma_v(k)$ is
a good indicator of the pairwise velocity dispersion
$\sigma_{PVD}(r)$, and can be measured in a redshift catalog (see
Section 4).

We have fitted the scaled damping function of our simulation with the
following form
\beq
D(k,\mu)={1\over 1+{1\over 2}(k\mu\sigma_v(k))^2+
\eta (k\mu\sigma_v(k))^4 }\,.
\label{fitting}
\eeq
The fitting values for the parameter $\eta$ are given in Table~2 for
each model. The above fitting formula describes our simulation results
very well, as shown by Figure 5. The formula can also be used to
compare the theoretical models with the statistic in future large
redshift surveys of galaxies.

It is quite useful to present a simple method for predicting
$P^S(k,\mu)$ in the theoretical models. One way is to combine
Eq.(\ref{fitting}) and Fig.~6. One could interpolate the data points
in Fig.~6 to get $\sigma_v(k)$ and use Eq.(\ref{fitting}) to predict
the damping function for the models. An alternative way is to replace
$\sigma_v(k)$ with the PVD of the {\it dark matter}
$\sigma^{dm}_{PVD}(1/k)$, since the dark matter PVD can be predicted
with the fitting formula given by \cite{mjb97}(1997). Figure~\ref{fig7}
shows the damping function when $\sigma^{dm}_{PVD}(1/k)$ is used for
$\sigma_v(k)$. The scatters in this figure are slightly larger than
those in Figure~\ref{fig5} as expected, but the scaling relation of
$D(k,\mu)$ still reasonably holds. The damping function could still
be described by the fitting formula (\ref{fitting}) with the fitting
values of $\eta$ listed in Table~3. One can use these results to
predict the redshift power spectrum $P^S(k,\mu)$.

We note that the fitting formula of this paper is valid only for the
models studied here. It is however possible to work out an analytical
model for the redshift power spectrum which is generally applicable to
CDM models and to the peak and cluster-weight bias schemes, based on
the known physical properties of dark matter halos (cf. \cite{mjb97} 1997;
\cite{ma2000}; \cite{mjw97}). We will study such an analytical model 
in a subsequent paper.

\section{Discussion and Conclusion}

As we have shown, there exist good scaling relations for the damping
function $D(k,\mu)$ when the scaling variable $k\mu\sigma_v(k)$ is
used. The conclusion is valid for all the galaxy formation models
examined in the paper (i.e. 3 CDM models and 2 bias models), although
the scaling form depends both on the dark matter models and the bias
models. $D(k\mu\sigma_v(k))$ falls with $k\mu\sigma_v(k)$ slightly
faster in the SCDM model than in the two low-density models. It also
falls faster in the cluster-over-weighted models (Peak models) than in
the CLW models.  For all the models, the damping function $D(k,\mu)$
is approximately described by the Lorentz form for large scales where
$D>0.1$, but the simulation result is below the analytical form for
smaller scales. The detailed form of the scaling relation is expected
to reflect the distribution function of the pairwise velocity as well
as the coupling between the velocity and density. Thus it depends on
the details of the galaxy formation model and the bias model. We have
presented an accurate fitting formula for the damping functions for
the models studied.

Our results have several important implications for observation.
Because the models studied cover a wide range of parameters, we
conjecture that a scaling relation of $D(k,\mu)$ also holds for the
galaxies in our Universe. This scaling relation can be measured by the
method used in this paper, by analyzing the redshift power spectrum
for a redshift catalog of galaxies. The resulting velocity dispersion,
though slightly different from the PVD defined by the 3-D peculiar
velocity, is a good indicator for it. The fact that both the scaling
form of $D(k\mu\sigma_v(k))$ and the velocity dispersion depend on the
dark matter model as well as on the bias model implies that these
quantities can become effective observational tests for theoretical
models. The determination for the scaling relation is also important
for determining the $\beta$ value when the redshift power spectrum can
be precisely measured only up to the scale $2\pi/k\approx 60 \mpc$. On
this scale or smaller, the nonlinear motion still has appreciable
effects and can be corrected by scaling the damping function.

In the traditional analysis of the two-point redshift correlation
function (TPRCF, \cite{1983ApJ...267..465D}), a functional form must
first be assumed for the DF of the pairwise velocity. The validity of
the functional form is checked by matching the model TPRCF, which is a
convolution of the the real-space two-point correlation with the DF of
the pairwise velocity, with the observational data. This method has
been applied to various redshift surveys (
\cite{1993MNRAS.264..825M}; \cite{1995AJ....110..477M}; 
\cite{1994MNRAS.267..927F}; \cite{ratcliffeetal1998}; 
\cite{postmanetal1998}; \cite{1999AJ....118.2561G};
\cite{1999ApJ...524...31S}). However, searching for the
functional form of the DF of the pairwise velocity in parameter space
is much more difficult than determining its Fourier counterpart, the
damping function, from an analysis of the redshift power spectrum.
Moreover, in modeling the TPRCF, an infall velocity as a function of
$r$ must be assumed (\cite{1983ApJ...267..465D}). The functional form of
the infall velocity is less well understood (\cite{1995AJ....110..477M};
\cite{jingboerner1998}) and much more difficult to
determine in observations than the the single parameter $\beta$ in the
power spectrum analysis. The velocity dispersions measured in both
analyses are shown to have similar accuracy (difference $\ls 20\%$)
compared to the 3-D peculiar velocity dispersion (e.g. JMB98;
\cite{jingboerner1998}).  Through this comparison, it is clear that
the two types of the redshift clustering analysis complement each
other, but an analysis of the redshift power spectrum has the above
mentioned advantages over the analysis of TPRCF.

In previous studies of the redshift power spectrum in CDM models
(e.g. \cite{1994MNRAS.267..785C}, 1995; \cite{2000ApJ...528...30M};
\cite{1997ApJ...475..414B}; \cite{1996MNRAS.279L...1F}; \cite{taylor};
\cite{1999MNRAS.310.1137H}), the authors 
studied the behavior of
$P^S(k,\mu)$ mainly in the quasi-linear regime, with emphasis on the
agreement of Equation (\ref{lorentz}) with the simulation data. Their
results were presented usually for the monopole and quadrupole of
$P^S(k,\mu)$ only, instead of the full dependence on $\mu$ as in this
paper. With the exception of Bromley et al. (1997) who considered dark
matter halos, most previous work was concerned with the redshift power
spectrum for the dark matter only.  In comparison, our present work
has focused on the full dependence of $P^S(k,\mu)$ on $\mu$ from the
quasilinear to the highly non-linear regime. To achieve this goal, we
have used a large set of high-resolution N-body simulations. We have
also paid close attention to the artificial biases introduced by the
FFT method, and we have included two distinctive bias models in order
to study the dependence of $P^S(k,\mu)$ on the galaxy bias.  Moreover,
we have given an accurate fitting formula for $P^S(k,\mu)$ for all the
models studied here.  Thus we present here a very thorough study of
the redshift power spectrum.

In summary, we have carried out a very systematic, detailed analysis
for the redshift distortion of the power spectrum in currently popular
models of galaxy formation, paying special attention to the strongly
non-linear regime. Three CDM models and two distinct bias models for
each CDM model are considered, with a total of nine models.  A large
set of high-resolution simulations of $256^3$ particles are used to
trace the linear and nonlinear clustering in these models. We have
carefully checked the FFT method and avoided those systematic biases
inherent to the method on small scale (or large $k$). Our main results
are:
\begin{enumerate}

\item The redshift power spectrum $P^S(k,\mu)$ can be accurately
  expressed by Equation (1) with the damping function $D(k,\mu)$ being
  a scaling function of the variable $k\mu \sigma_v(k)$ in all the
  models studied here. An accurate fitting formula has been given for
  these scaling functions.
\item A mild variation of $\sigma_v(k)$ with the scale $k$ is required
  to improve the scaling relation of $D(k,\mu)$ among different
  scales. The behavior of $\sigma_v(k)$ with the scale is
  qualitatively similar to that seen in the PVD defined by the 3D
  peculiar velocity. Therefore, the variation of $\sigma_v(k)$ with
  the scale in fact reflects the separation dependence of the PVD,
  although there is about $20\%$ systematic difference between these
  two quantities.
\item The damping function is described well by the Lorentz form for
  $D(k\mu \sigma_v)>0.1$ but falls faster than the Lorentz form on
  smaller scales ($D(k\mu \sigma_v)<0.1$).  The most likely cause for
  the deviation is the strong coupling between the nonlinear motion
  and the small scale structures.
\item The functional form of $D(k\mu \sigma_v)$ as well 
as the quantity $\sigma_v(k)$ depends on the dark matter models and
the bias recipes. An observational measurement of these two quantities
can serve as an interesting test for models of galaxy formation.

\end{enumerate}

Because of these interesting features of $P^S(k,\mu)$, an
observational analysis of the redshift power spectrum, though it is
largely complementary to, has several obvious advantages over the
traditional analysis of the two-point correlation function. In a
subsequent paper, we (\cite{jingboerner2000}) will apply the results
of this paper to measure the damping function and the velocity
dispersion for the currently largest redshift survey- the Las Campanas
Redshift Survey (\cite{shectmanetal1996}).

All the data presented in the figures are available to the interested
readers in electronic form upon request.

\acknowledgments 

We are grateful to Yasushi Suto for the hospitality extended to us at
the physics department of Tokyo university where most of the
computation was completed. We thank the referee for helpful comments
which improve the presentation of this paper. J.Y.P. gratefully
acknowledges the receipt of a NAO COE foreign research fellowship. The
work is supported by the One-Hundred-Talent Program and by The Major
State Basic Research Development Program to Y.P.J., and by SFB375 to
G.B.. The simulations were carried out on VPP/16R and VX/4R at the
Astronomical Data Analysis Center of the National Astronomical
Observatory, Japan.

\newpage
\begin{table}[\begin{table}[h]
\begin{center}
  Table~1.\hspace{4pt} Model parameters \\ 
\end{center}
\vspace{6pt}
\begin{center}
\begin{tabular}{ccccccc}
\hline\hline\\[-6pt]
Model & $\Omega_0$ &  $\lambda_0$  
&  $\Gamma$ &   $\sigma_8$ \\ 
[4pt]\hline \\[-6pt]
SCDM & 1.0  & 0.0 & 0.5 & 0.6 \\
OCDM & 0.3  & 0.0 & 0.25 & 1.0\\
LCDM & 0.3  & 0.7 & 0.21 & 1.0\\
\hline
\end{tabular}
\end{center}
\end{table}
\begin{table}[\begin{table}[h]
\begin{center}
  Table~2.\hspace{4pt} The fitting values of $\eta$ \\ 
\end{center}
\vspace{6pt}
\begin{center}
\begin{tabular}{ccccccc}
\hline\hline\\[-6pt]
 & DM &  Peaks  
&  CLW \\ 
[4pt]\hline \\[-6pt]
SCDM & 0.00965& 0.0204& 0.00792 \\
OCDM & 0.00330& 0.0168&0.00171 \\
LCDM & 0.00309& 0.0168& 0.00145\\
\hline
\end{tabular}
\end{center}
\end{table}
\newpage
\begin{table}[\begin{table}[h]
\begin{center}
  Table~3.\hspace{4pt} The fitting values of $\eta$ when $\sigma_v(k)$ is replaced with $\sigma_{PVD}(1/k)$ \\ 
\end{center}
\vspace{6pt}
\begin{center}
\begin{tabular}{ccccccc}
\hline\hline\\[-6pt]
 & DM &  Peaks  
&  CLW \\ 
[4pt]\hline \\[-6pt]
SCDM & 0.01118& 0.0401 & 0.00327 \\
LCDM & 0.00759& 0.0481 & 0.00089 \\
OCDM & 0.00566& 0.0429 & 0.00021 \\
\hline
\end{tabular}
\end{center}
\end{table}
\newpage

\begin{figure}
\epsscale{1.0} \plotone{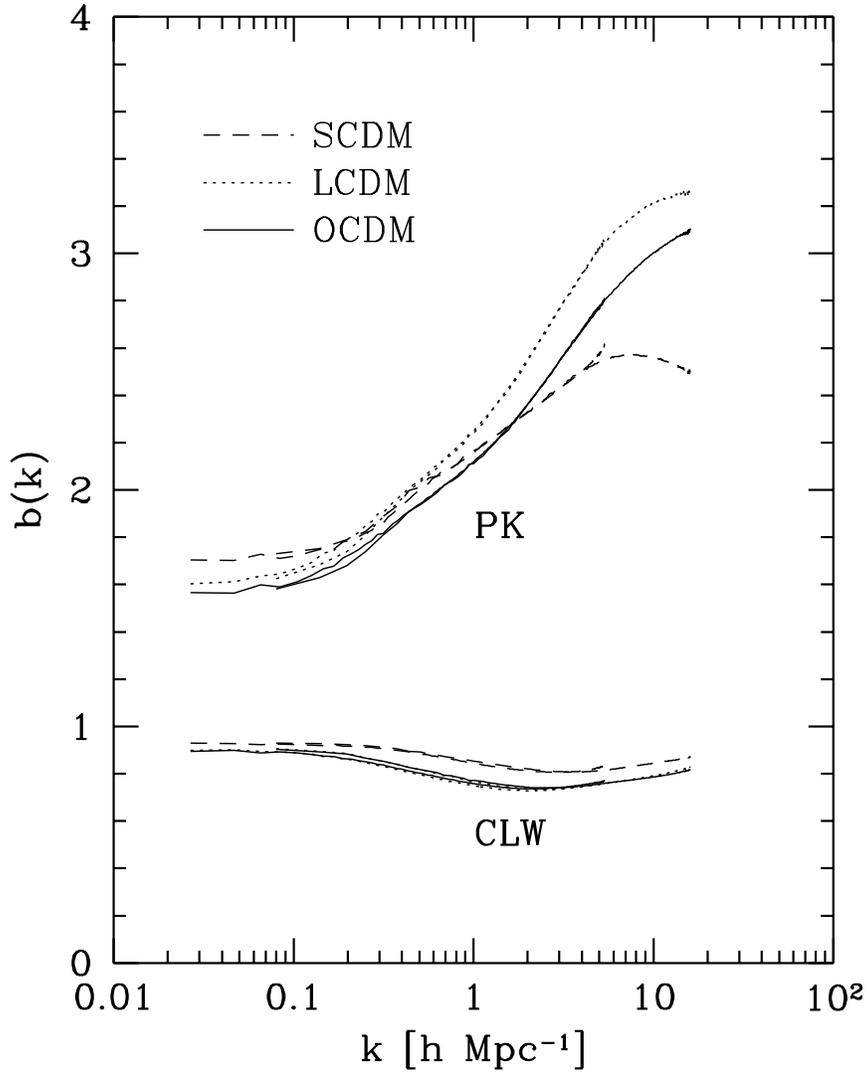}
\caption{The bias factor $b(k)$ measured as the square root of the 
ratio of the power spectrum of the biased tracer to that of the 
dark matter.}\label{fig1}\end{figure}

\begin{figure}
\epsscale{1.0} \plotone{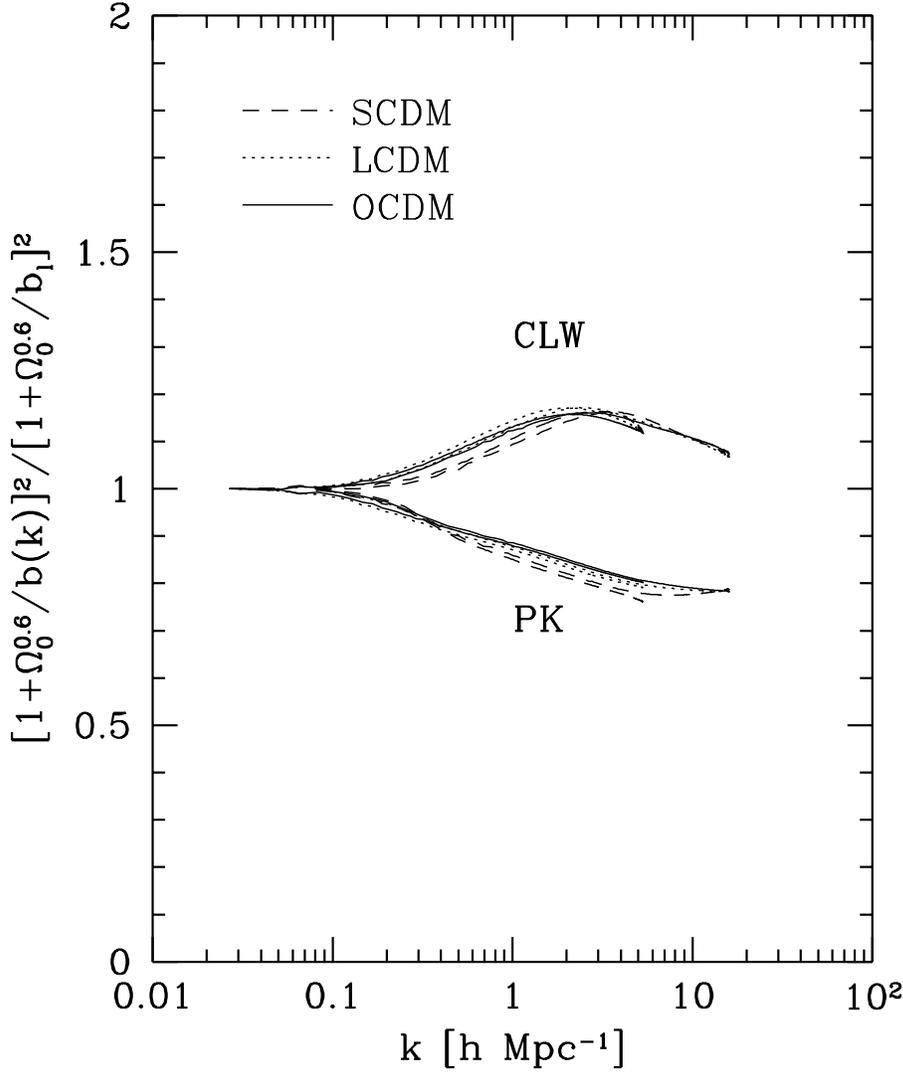}
\caption{The
ratio of $[1+\Omega_0^{0.6}/b(k)]^2$ to $[1+\Omega_0^{0.6}/b_{l}]^2$,
where $b_{l}$ is the bias on the linear scale. The curves represent
the maximum effect that the non-linear bias has on the linear Kaiser
effect at $\mu=1$. }\label{fig2}\end{figure}

\begin{figure}
\epsscale{1.0} \plotone{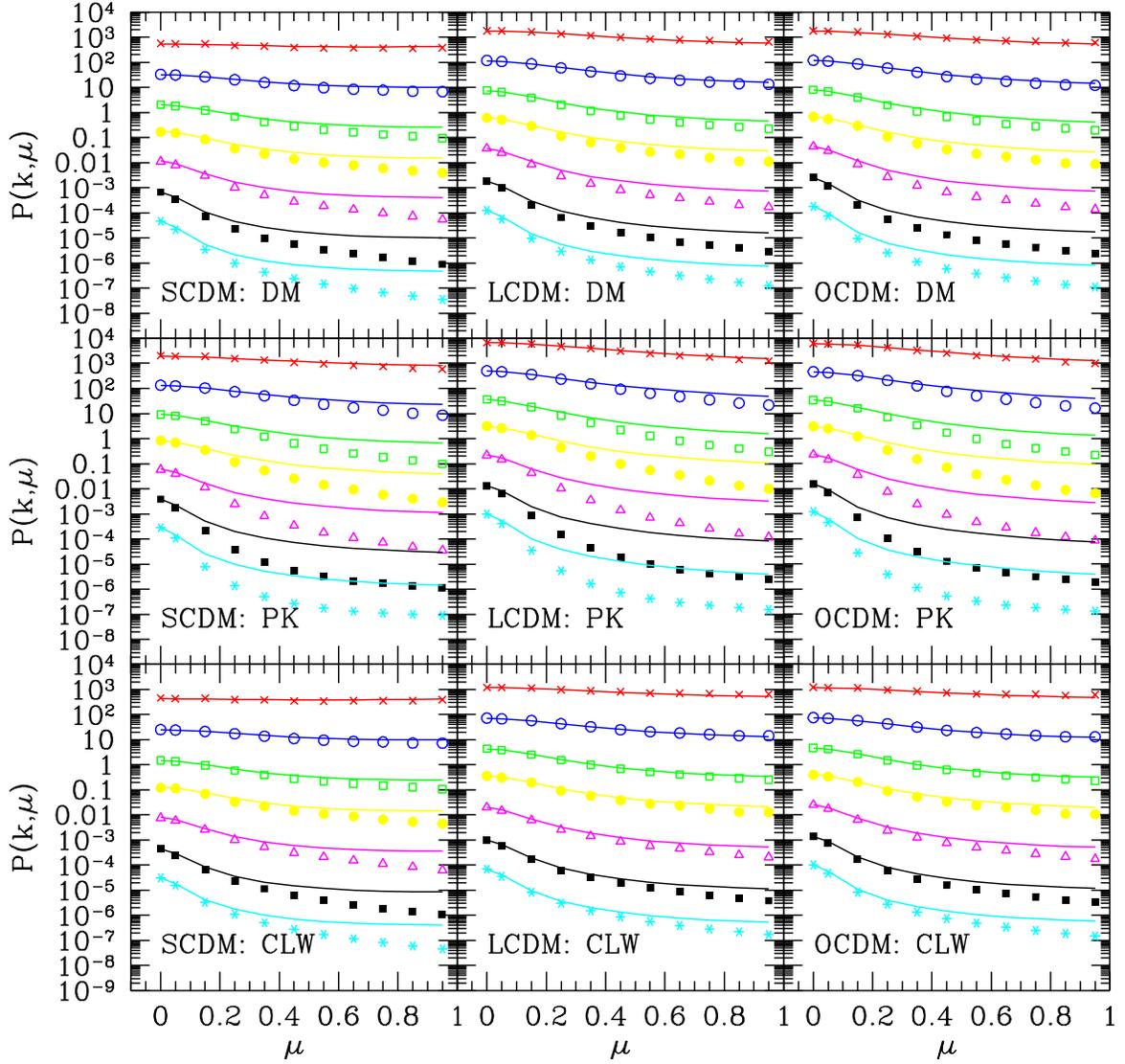}
\caption{
  The redshift power spectrum $P^S(k,\mu)$ determined for the
  different tracers of the three CDM models. The top panels are for
  the dark matter (DM), the middle for the peaks (PK), and the bottom
  for the cluster under weighted population (CLW). In each panel,
  $P^S(k,\mu)$ at different $k$ are plotted with different
  symbols. From the top (crosses) to the bottom (stars), the values of
  $k$ are 0.35, 0.56, 0.90, 1.1, 1.7, 2.7, 3.4 $\mpc$ respectively,
  and the $P^S(k,\mu)$ are multiplied by $1$, $10^{-1}$, $10^{-2}$,
  ..., $10^{-6}$ respectively for clarity. The lines are the model
  prediction of Equation (\ref{lorentz}) where the Lorentz form is
  used for the damping function.  }\label{fig3}\end{figure}

\begin{figure}
\epsscale{1.0} \plotone{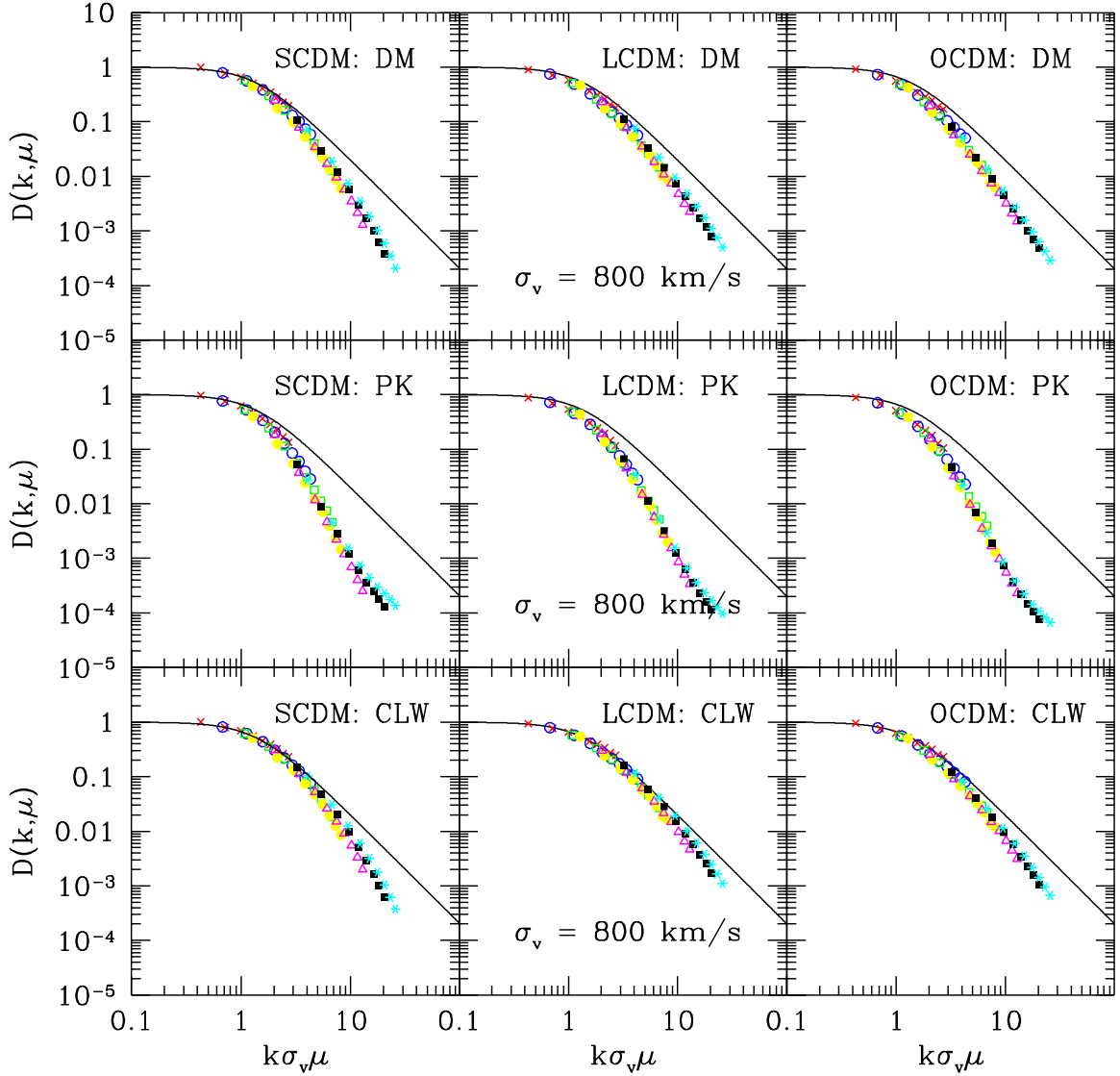}
\caption{
  The damping function $D(k,\mu)$ determined according to Equation
  (\ref{dd}) is plotted as a function of $k\mu\sigma_v$ with
  $\sigma_v$ fixed to $800 \kms$. The curves are the prediction of the
  Lorentz form.  The symbols and the labels are the same as in Fig.~2
  }\label{fig4}\end{figure}

\begin{figure}
\epsscale{1.0} \plotone{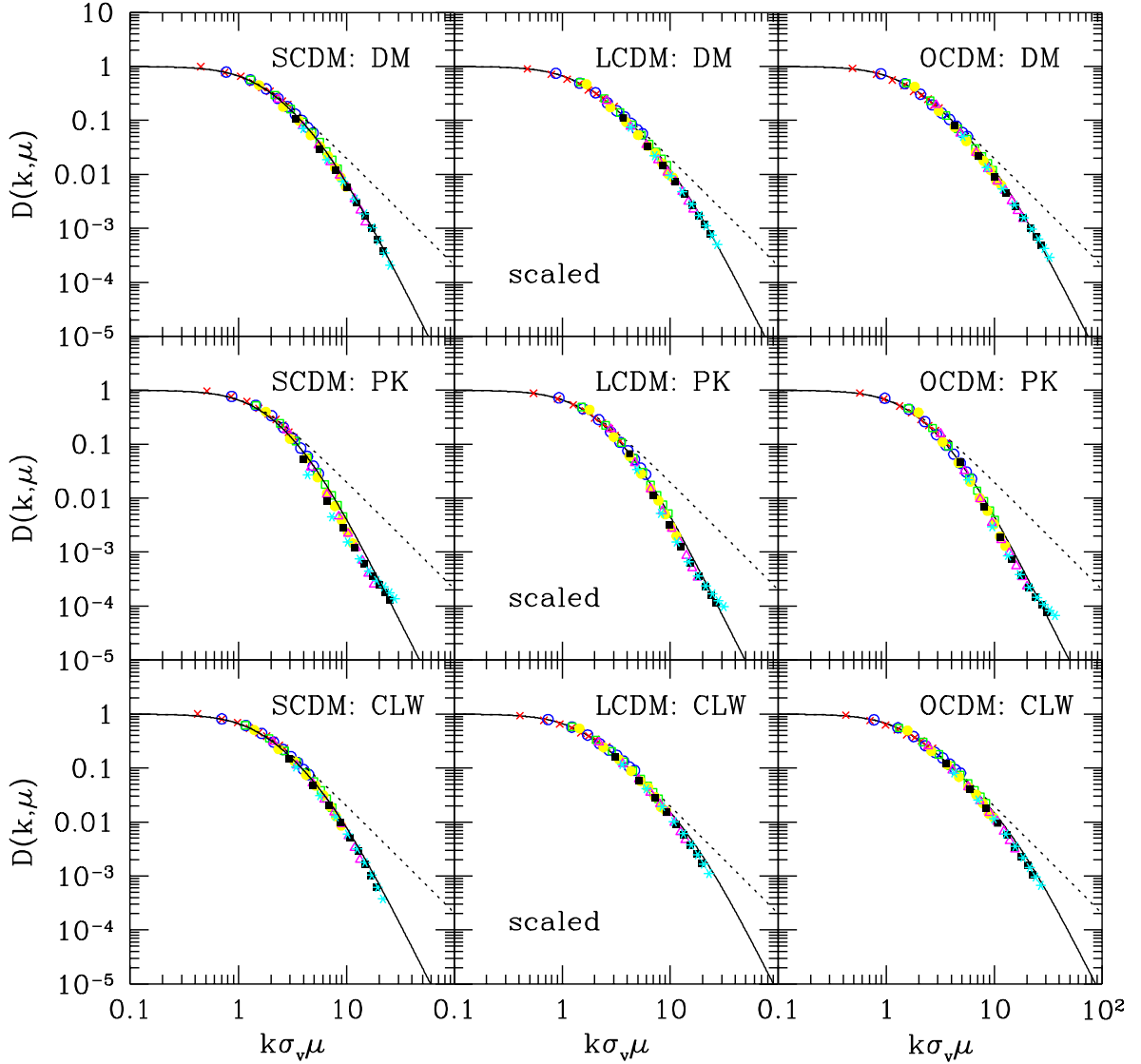}
\caption{
  The same as Fig.~4, but the velocity dispersion $\sigma_v$ is
  allowed to vary with the wavenumber $k$. The dashed curves are the
  prediction of the Lorentz form and the solid ones are given by the
  fitting formula Eq.(\ref{fitting}) with the
  $\eta$ given in Table 2. The values of $\sigma_v(k)$ used
  for this plot are given in Fig.~6.  }
\label{fig5}\end{figure}

\begin{figure}
\epsscale{1.0} \plotone{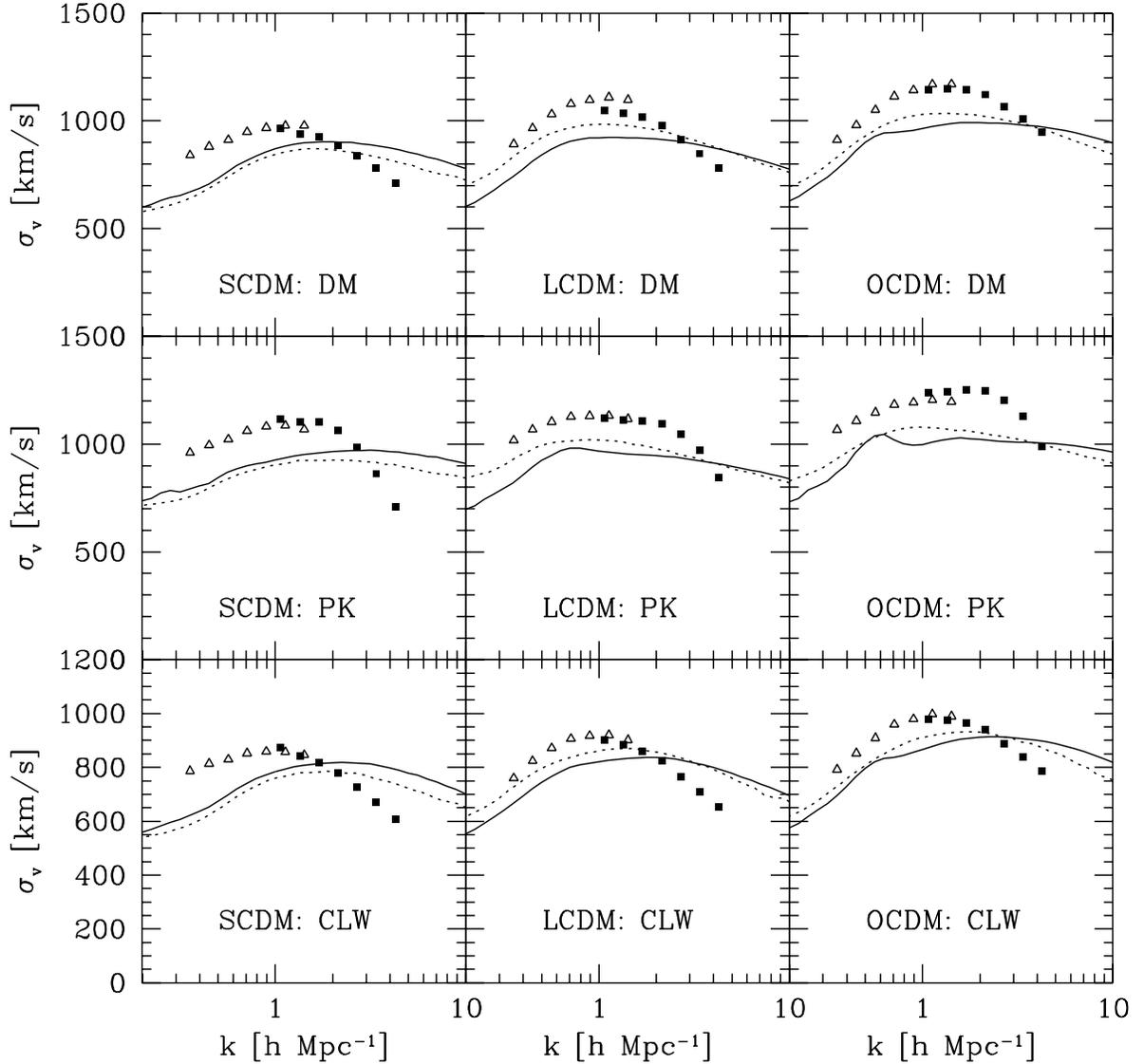}
\caption{The velocity dispersion $\sigma_v (k)$ (symbols) 
required to make the
damping function a scaling function of $k\mu\sigma_v$, compared to the
pairwise velocity dispersions measured from the true peculiar
velocity(curves). The open symbols and the dotted lines are for the
simulations of $300\mpc$, and the solid symbols and the solid lines
for those of $100\mpc$. The systematic difference of the velocity
dispersion at $k\sim 1\mpci$ between the two box sizes, which is small
(5\% only), is due to the different long wavelength cutoff in the
simulation. }\label{fig6}\end{figure}

\begin{figure}
\epsscale{1.0} \plotone{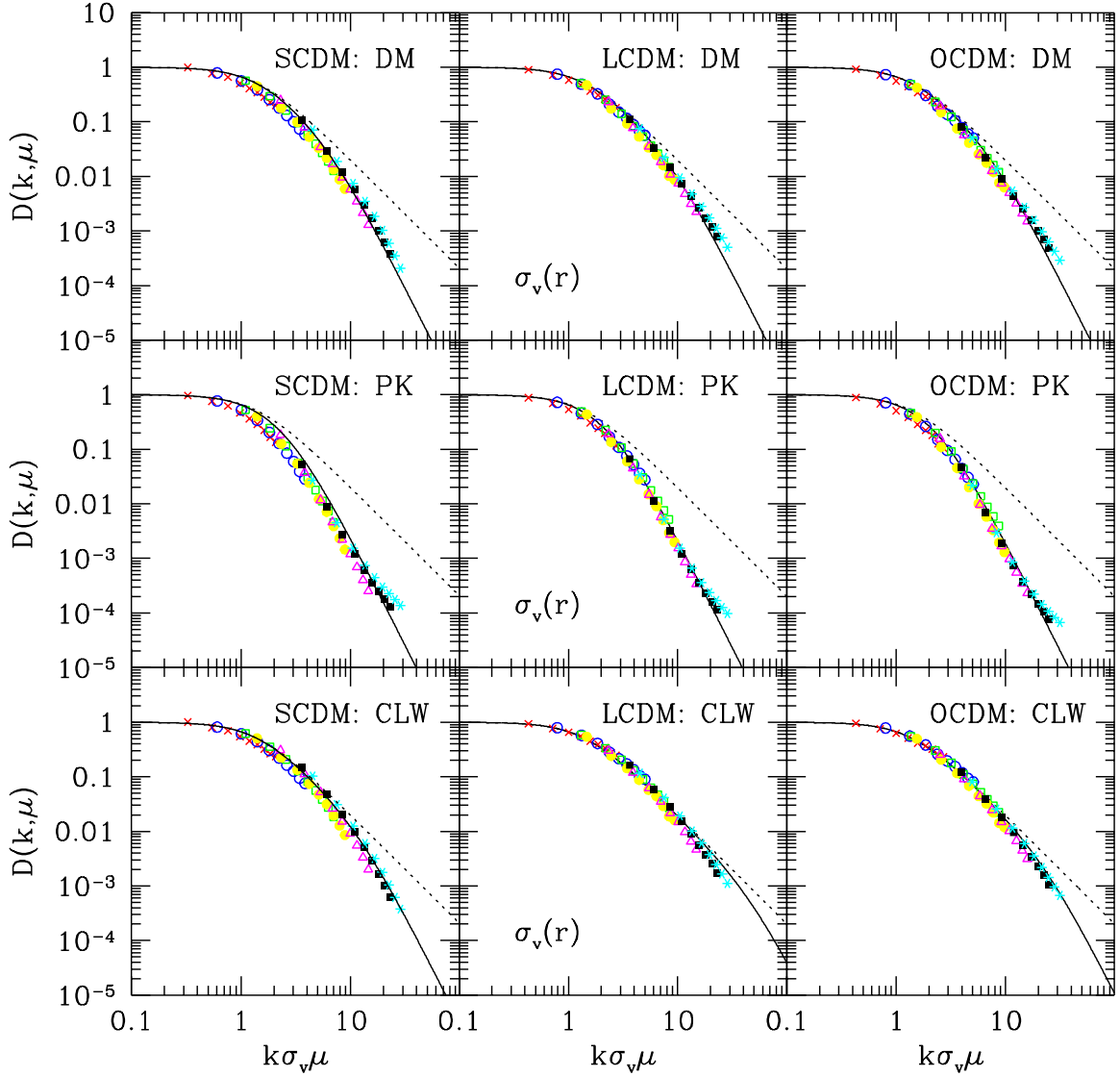}
\caption{
  The same as Fig.~5, but the velocity dispersion $\sigma_v(k)$ is
taken from the pairwise velocity dispersion $\sigma_{PVD}(r)$ at $r=1/k$. The
dashed curves are the prediction of the Lorentz form and the solid
ones are given by the fitting formula Eq.(\ref{fitting}) with the
$\eta$ given in Table 3.}
\label{fig7}\end{figure}


\begin{thebibliography}{}

\bibitem[Adelberger, et al. 1998]{adelbergeretal1998} Adelberger, K. L., Steidel, C. C., Giavalisco, M. , Dickinson, M. , Pettini, M.  \& Kellogg, M.  1998, \apj, 505, 18 
\bibitem[Alcock \& Paczynski 1979]{1979Natur.281..358A} Alcock, C. \& Paczynski, B. 1979, \nat, 281, 358 
\bibitem[Ballinger, Peacock \& Heavens 1996]{1996MNRAS.282..877B} Ballinger, W. E., Peacock, J. A. \& Heavens, A. F. 1996, \mnras, 282,877 
\bibitem[Bardeen, et al. 1986]{1986ApJ...304...15B} Bardeen, J. M., Bond, J. R., Kaiser, N. \& Szalay, A. S. 1986, \apj, 304, 15 
\bibitem[Benson, et al. 1999]{bensonetal1999} Benson, A. J., Baugh, C. M., Cole, S., Frenk, C. S., \& Lacey, C. G. 1999, astro-ph/9911179
\bibitem[Bromley, Warren \& Zurek 1997]{1997ApJ...475..414B} Bromley, B. C., Warren, M. S. \& Zurek, W. H. 1997, \apj, 475, 414 
\bibitem[Carlberg, et al. 1996]{1996ApJ...462...32C} Carlberg, R. G., Yee, H. K. C., Ellingson, E., Abraham, R., Gravel, P., Morris, S. \& Pritchet, C. J. 1996, \apj, 462, 32 
\bibitem[Cole, Fisher \& Weinberg 1994]{1994MNRAS.267..785C} Cole, S., Fisher, K. B. \& Weinberg, D. H. 1994, \mnras, 267, 785 
\bibitem[Cole, Fisher \& Weinberg 1995]{1995MNRAS.275..515C} Cole, S. , Fisher, K. B. \& Weinberg, D. H. 1995, \mnras, 275, 515 
\bibitem[Davis \& Peebles 1983]{1983ApJ...267..465D} Davis, M. \& Peebles, P. J. E. 1983, \apj, 267, 465 
\bibitem[Diaferio \& Geller 1996]{1996ApJ...467...19D} Diaferio, A.  \& Geller, M. J. 1996, \apj, 467, 19 
\bibitem[Efstathiou, et al. 1988]{1988MNRAS.235..715E} Efstathiou, G. , Frenk, C. S., White, S. D. M. \& Davis, M.  1988, \mnras, 
\bibitem[Fisher \& Nusser 1996]{1996MNRAS.279L...1F} Fisher, K. B. \& Nusser, A.  1996, \mnras, 279, L1 
\bibitem[Fisher, et al. 1994]{1994MNRAS.267..927F} Fisher, K. B., Davis, M., Strauss, M. A., Yahil, A. \& Huchra, J. P. 1994, \mnras, 267, 927 
\bibitem[Geller \& Peebles 1973]{1973ApJ...184..329G} Geller, M. J. \& Peebles, P. J. E. 1973, \apj, 184, 329
\bibitem[Grogin \& Geller 1999]{1999AJ....118.2561G} Grogin, N. A. \& Geller, M. J. 1999, \aj, 118, 2561 
\bibitem[Hatton \& Cole 1999]{1999MNRAS.310.1137H} Hatton, S.  \& Cole, S.  1999, \mnras, 310, 1137 
\bibitem[Jing (1992)]{jing1992} Jing, Y. P. 1992, ph.D.thesis, SISSA, Trieste
\bibitem[Jing (1998)]{jing1998} Jing, Y.P., 1998, ApJ, 503, L9
\bibitem[Jing (2000)]{jing2000} Jing, Y.P. 2000, in preparation
\bibitem[Jing \& B\"orner 1998]{jingboerner1998} Jing, Y. P. \& B\"orner, G. 1998, \apj, 503, 37 
\bibitem[Jing \& B\"orner 2000]{jingboerner2000}Jing, Y. P. \& B\"orner, G. 2000,
\apj, (submitted)
\bibitem[Jing \& Suto (1998)]{1998ApJ...494L...5J} Jing, Y. P. \& Suto, Y.  1998, \apjl, 494, L5 
\bibitem[Jing, Mo \& B\"orner 1998]{1998ApJ...494....1J} Jing, Y. P., Mo, H. J. \& B\"orner, G. 1998, \apj, 494, 1 
\bibitem[Jing, et al. 1994]{1994A&A...284..703J} Jing, Y. P., Mo, H. J., B\"orner, G. \& Fang, L. Z. 1994, \aap, 284, 703
\bibitem[Juszkiewicz, Fisher \& Szapudi 1998]{1998ApJ...504L...1J} Juszkiewicz, R. , Fisher, K. B. \& Szapudi, I.  1998, \apjl, 504, L1
\bibitem[Kaiser 1987]{1987MNRAS.227....1K} Kaiser, N.  1987, \mnras, 227, 1 
\bibitem[Ma \& Fry 2000]{ma2000} Ma, C.\ and Fry, J.\ N.\ 
2000, \apjl, 531, L87
\bibitem[Magira, Jing \& Suto 2000]{2000ApJ...528...30M} Magira, H. , Jing, Y. P. \& Suto, Y.  2000, \apj, 528, 30 
\bibitem[Marzke, et al. 1995]{1995AJ....110..477M} Marzke, R. O., Geller, M. J., da Costa, L. N. \& Huchra, J. P. 1995, \aj, 110, 477 
\bibitem[Matsubara 1999]{matsubara_bias} Matsubara, T.\ 1999, \apj, 
525, 543 
\bibitem[Matsubara \& Suto 1996]{1996ApJ...470L...1M} Matsubara, T.  \& Suto, Y.  1996, \apjl, 470, L1 
\bibitem[Matsubara, Szalay and Landy (2000)]{matsubaraetal2000} 
Matsubara, T., Szalay, A. S. and Landy, S. D. 2000, \apjl, 535, L1
\bibitem[Mo, Jing \& B\"orner 1993]{1993MNRAS.264..825M} Mo, H. J., Jing, Y. P. \& B\"orner, G. 1993, \mnras, 264, 825
\bibitem[Mo, Jing \& White 1997]{mjw97} Mo, H.\ J., Jing, 
Y.\ P.\ and White, S.\ D.\ M.\ 1997, \mnras, 284, 189 
\bibitem[Mo, Jing \& B\"orner]{mjb97} Mo, H.\ J., Jing, 
Y.\ P.\ and B\"orner, G.\ 1997, \mnras, 286, 979 
\bibitem[Peacock \& Dodds 1994]{1994MNRAS.267.1020P} Peacock, J. A. \& Dodds, S. J. 1994, \mnras, 267, 1020 
\bibitem[Peebles 1976]{1976Ap&SS..45....3P} Peebles, P. J. E. 1976, \apss, 45, 3 
\bibitem[Press \& Schechter (1974)]{1974ApJ...187..425P} Press, W. H. \& Schechter, P.  1974, \apj, 187, 425 
\bibitem[Postman et al. 1998]{postmanetal1998} 
Postman, M., Lauer, T. R., Szapudi, I. ;. and Oegerle, W. 1998, \apj, 506, 
33 
\bibitem[Seto \& Yokoyama 1998]{1998ApJ...492..421S} Seto, N.  \& Yokoyama, J. 'i.  1998, \apj, 492, 421 
\bibitem[Shectman et al. 1996] {shectmanetal1996} Shectman S. A., Landy, S.D.,
Oemler A., Tucker D.L., Lin H., Kirshner R.P., Schechter P.L., 1996,
ApJ, 470, 172
\bibitem[Sheth 1996]{1996MNRAS.279.1310S} Sheth, R. K. 1996, \mnras, 279, 1310 
\bibitem[Small, et al. 1999]{1999ApJ...524...31S} Small, T. A., Ma, C. -P. , Sargent, W. L. W. \& Hamilton, D.  1999, \apj, 524,31
\bibitem[Suto, et al. 1999]{1999PThPS.133..183S} Suto, Y., Magira, H., Jing, Y. P., Matsubara, T. \& Yamamoto, K. 1999, Progress in Theoretical Physics Supplement, 133, 183
\bibitem[Steidel et al. 1998]{steideletal1997} Steidel, C.C., Adelberger, K.L., Dickinson,
  M., Giavalisco, M., Pettini, M., Kellogg, M., 1998, ApJ, 492, 428
\bibitem[Ratcliffe et al. 1998]{ratcliffeetal1998} 
Ratcliffe, A., Shanks, T., Parker, Q. A. and Fong, R. 1998, \mnras, 296, 
191 
\bibitem[Taylor and Hamilton 1996]{taylor} Taylor, A. N. and 
Hamilton, A. J. S. 1996, \mnras, 282, 767 
\bibitem[Turner 1976]{1976ApJ...208...20T} Turner, E. L. 1976, \apj, 208, 20 
\bibitem[White, et al. (1987)]{1987ApJ...313..505W} White, S. D. M., Frenk, C. S., Davis, M.  \& Efstathiou, G.  1987, \apj, 313, 505 

\end{thebibliography}
\end{document}